\documentclass{aastex631}
\usepackage{graphicx}

\begin{document}

\title{Which Solar Latitude Follows the Sunspot Cycle Exactly?}

\author[0000-0002-6648-4170]{K P Raju}
\affiliation{Indian Institute of Astrophysics \\
Bangalore 560034 \\
India}



\begin{abstract}

The large-scale convection in the Sun known as supergranulation is manifested as a network structure 
on the solar surface. The network cells have an average lifetime of 24 hr, a size of about 30 Mm, and a lane width of about 6 Mm. We have obtained the lane widths and intensities at different latitudes from 
the Ca {\sc ii} K spectroheliograms from the 100 yr Kodaikanal archival data. We have then calculated the
cross correlation function of lane widths and intensities with sunspot number at every latitude from
60$^{\circ}$ N to 60$^{\circ}$ S. The correlation coefficients of the quantities show an approximate North-South symmetry with broad peaks around $\pm$(11--22)$^{\circ}$ latitude with values of about 0.8. The results imply that these latitudes follow the sunspot cycle strongly. The maximum correlation for the lane widths occurs (18$\pm$2)$^{\circ}$ N and (20$\pm$2)$^{\circ}$ S with no phase difference. For intensities, this happens at 
(13$\pm$2)$^{\circ}$ N and (14$\pm$2)$^{\circ}$ S with a phase difference of 1.25 to 1.5 yr. It is interesting to note that the lane width correlations peak during the solar maximum whereas the intensitiy correlations peak 1.25--1.5 yr after the solar maximum. The results, generally show that no unique latitude exactly follows the solar cycle for all quantities.
The results are important in flux transport on the solar surface and have implications for the quiet Sun UV irradiance variations.

\end{abstract}

\section{Introduction} \label{sec:intro}

The two basic scales of solar convection are the granulation and the supergranulation. Granulation is the small scale and the supergranulation is the large scale which is manifested as a network structure on the solar surface. The network cells have an average lifetime of 24 hr, a size of about 30 Mm, and a lane width of about 6 Mm. The network arises due to magnetic flux concentration at cell boundaries as a consequence of supergranular convection \citep{1962ApJ...135..474L,1964ApJ...140.1120S}. Skylab observations in the 1970s have shown that the chromospheric network extends to the transition region as the EUV network \citep{1976SoPh...46...53R}. The network is dominant in the mid-transition region, and it disintegrates in the corona \citep{1998A&A...335..733G}.
Several interesting findings in recent years have raised questions on the origin of
supergranulation and why they occur primarily at about 30 Mm scale.
The original suggestion by \citet{1964ApJ...140.1120S} that the Helium recombination plays a role in forming the supergranular scales is not supported by simulations or models \citep{2014ApJ...793...24L,2023SSRv..219...77H}.
A thermal convective origin of supergranulation has long been the credible explanation, but the contribution of other factors like rotation, magnetic field, and multi-scale convection is also being considered \citep{2018LRSP...15....6R}. 
\citet{1980SoPh...66..213D} and \citet{1990ApJ...351..309S} found that supergranulation
rotates faster than the surrounding plasma, which is referred to as superrotation.
\citet{2003Natur.421...43G} found that the supergranulation pattern had wave-like properties with
a typical period of 6–9 days, much longer than the individual supergranules' lifetime.
\citet{2010ARA&A..48..289G} use local helioseismology to show the effect of the Coriolis force on supergranular flows.
\citet{2014A&A...567A.138R} reported that supergranular motions reflect solar differential rotation and a poleward meridional flow. Also, the presence of magnetic fields was found to weaken diverging flows.
\citet{2016ApJ...829L..17C} suggest that the supergranular scale as the largest buoyantly driven mode of convection in the Sun.
\citet{2022ApJ...937...41G} found that the horizontal flow scales increase rapidly with depth. The total power of the convective flows is anticorrelated with the sunspot number variation in subsurface layers, and positively correlated at larger depths.
\citet{2024NatAs...8.1088H} analyzed Dopplergrams from SDO to identify 23,000
supergranules. They found that the vertical flows peak at a depth of 10,000 km, and the supergranular convection is not explained by mixing-length theory.

Solar activity levels may influence supergranulation properties such as its scale, lifetime, and flow velocity \citep{2007A&A...466.1123M,2008A&A...488.1109M}. The dependence of supergranular size on
the solar cycle has remained controversial. A decrease of cell sizes has been reported by \citet{1981SoPh...71..161S}, \citet{1999A&A...344..965B}, \citet{2002SoPh..207...11R}, \citet{1994SoPh..152..139K}, \citet{2004ApJ...616.1242D}, and \citet{2012ApJ...759..106H}, but \citet{1988SoPh..117..343W}, \citet{1989A&A...213..431M}, and \citet{2003A&A...405.1107M} pointed out an increase
of network cell sizes. \citet{2007A&A...466.1123M} suggest that a negative or a positive
correlation can be obtained, depending on whether the level of magnetic activity
is defined with respect to internetwork or network fields.
 \citet{2011ApJ...730L...3M} found a reduction of 0.5 Mm in the average supergranular radius during the cycle 23/24 minimum compared to the cycle 22/23 minimum.
\citet{2017ApJ...844...24M} found that mean scale values are highly correlated with
the sunspot cycle amplitude.
\citet{2017ApJ...841...70C} showed that the active region supergranule mean scale
varies in phase with the solar cycle, whereas for the quiet region
supergranule mean scale it is the opposite.

In addition to the size, there is another length scale associated with the supergranulation: the lane width. In their seminal work, Simon \& Leighton (1964) obtained the lane width through a manual autocorrelation method and interpreted it as a measure of magnetic flux at the supergranular cell boundaries. Although the lane width and size show some similarities in their behavior \citep{1970SoPh...13..292S,2003A&A...408..743G}, their interrelationship and the possible role of lane width in supergranular origin are not known.
However, several interesting results have been obtained from lane width studies. 
\citet{1970SoPh...13..292S} from Ca K spectroheliogram measurements reported asymmetry in the supergranular length scales. Both lane widths and sizes were found to be lengthened in the direction of the solar rotation. The lane widths were found to have an almost constant width up to the upper transition region and then fan out rapidly at coronal temperature which supports the funnel model of network magnetic field \citep{1999ApJ...522..540P}. \citet{2003A&A...408..743G} found that the size and lane widths generally increase with the formation temperatures of the chromospheric and transition region emission lines.
From SOHO/SUMER data, \citet{2008A&A...482..267T} found that the network lane width is smaller in the chromosphere than in the transition region.
In an earlier work \citep{2016SoPh..291.3519R}, we have obtained the network lane width from the SOlar and Heliospheric Observatory (SOHO)/Coronal Diagnostic Spectrometer (CDS) synoptic images of the Sun in 
He {\sc i} 586 {\AA} and O {\sc v} 630 {\AA}. The 
lane widths are found to be correlated with the solar cycle variation with a lag of about ten months. The data also show large asymmetry in network lane widths in the horizontal and vertical directions caused by image distortions in the CDS due to instrumental effects.
We have also obtained the network lane widths and intensities from the Ca II K spectroheliograms from the Kodaikanal archival data \citep{2018MNRAS.478.5056R}. 
The results show that both quantities are dependent on
the solar cycle. Also, a varying phase difference between the quantities has been noticed in different
solar latitudes, and evidence of equator-ward flux transfer.
It is also found that the lane widths, obtained near the mid-latitudes during the sunspot cycle
minima, are strongly correlated to the following sunspot number maxima.  The
strong correlation of the two parameters provides a simple way to predict the maximum
sunspot number about 4-5 yr in advance \citep{2023ApJ...959L..24R}.

As the previous paragraph shows, the lane width and
intensities depend on the solar cycle. The lane width from SOHO/CDS EUV images shows a
time lag of about 10 months in the equatorial region. In the present paper, we examine the
relationship between the solar cycle and two physical quantities, lane widths and intensities,
in more detail. We would like to see how the time lag between the quantities varies in different
latitudes. We want to examine the cross correlation between these quantities and see where
the correlation coefficient reaches a maximum. We examine all latitudes from 60$^{\circ}$ N to 60$^{\circ}$ S
to see whether any particular latitude follows the solar cycle exactly. We
use the newly calibrated Kodaikanal archival data \citep{2014SoPh..289..137P} in the
study. The following sections describe the details of the data analysis, results, and conclusions.

\begin{figure}
\includegraphics[width=15cm,height=23cm]{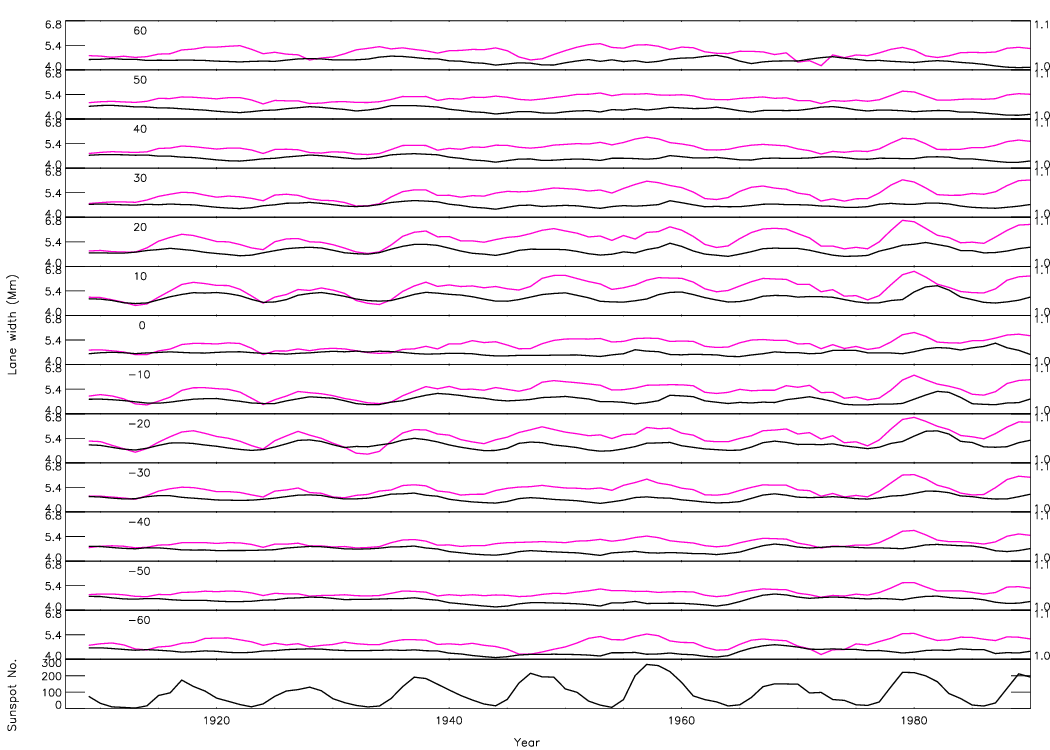}
\caption{Yearly averaged supergranular lane widths (pink) and Ca K intensity (black) as a function of time for different latitudes from 60$^{\circ}$ N to 60$^{\circ}$ S with an interval of 10$^{\circ}$. The y-axis on the right gives the normalized intensity. The bottom panel shows the variation of the yearly averaged sunspot number with time. Sunspot data from the World Data Center SILSO, Royal Observatory of Belgium, Brussels.
\label{fig:general}}
\end{figure}

\section{Data \& Analysis} \label{sec:data}

The present analysis uses about 34000 Ca {\sc ii} K spectroheliograms from the Kodaikanal Solar Observatory.  
The instrument used is the same throughout the 100 years of observation. It is a
spectroheliograph where no filter is involved. The exit slit is centered at the
Ca-K line at 3933.67 {\AA} with a spectral window of 0.5 \AA. The centering of the
Ca-K line has a maximum uncertainty of about 0.1 {\AA} due to the visual setting of the
spectrum and due to the stability of the spectroheliograph \citep{2014SoPh..289..137P}.
This affects the contrast of the spectroheliograms. There are also other factors
affecting the contrast, such as change of photographic emulsions, development,
sky transparency, and variations in density-to-intensity conversion. The contrast
changes are effectively corrected using 'equal-contrast technique' \citep{2021ApJ...908..210S}.
The images are corrected for limb darkening, and their contrast is adjusted until the FWHM of the intensity distribution attains a value between 0.10 and 0.11. The resulting intensity is normalized with values varying from 1.0 to 1.1 for the quiet Sun, 1.1 to 1.2 for the active network, and $>$ 1.2 for active region plages.
 The spatial resolution of the data is about 2${\arcsec}$ \citep{1967SoPh....1..151B}. The data is obtained
 from 1907 to 2007 but we have considered images only till 1990 as there was a seeing deterioration after this time \citep{sri17}.
The details of the data analysis are described in \citep{2020ApJ...899L..35R}. The image windows of size 120 (arcsec)$^{2}$ contain about 16 supergranules are taken from the central meridian in the latitude range 60$^{\circ}$ N to 60$^{\circ}$ S with an interval of 1$^{\circ}$. The window sizes are corrected for foreshortening effects on the solar surface. 
Active region windows are avoided by applying the following intensity criterion.
Based on the mean intensity of the windows, we rejected those at the
top and bottom 5 \% of the distribution \citep{2023ApJ...959L..24R}. This removes the low exposure regions
at the lower end and the active regions and active network at the higher end. The
threshold is decided through a trial and error method. Varying the threshold
by a few percentage does not significantly affect the correlation.
Lane width is measured as the width of the autocorrelation function \citep {2023ApJ...959L..24R} 
of the image window.
Lane widths and the mean intensity are obtained for every image window and the annual variations are removed by averaging them over a year. 
The corrected lane widths and intensities are obtained as functions of latitude and time. 

\begin{figure}
\centering
{\includegraphics[width=10cm,height=18cm]{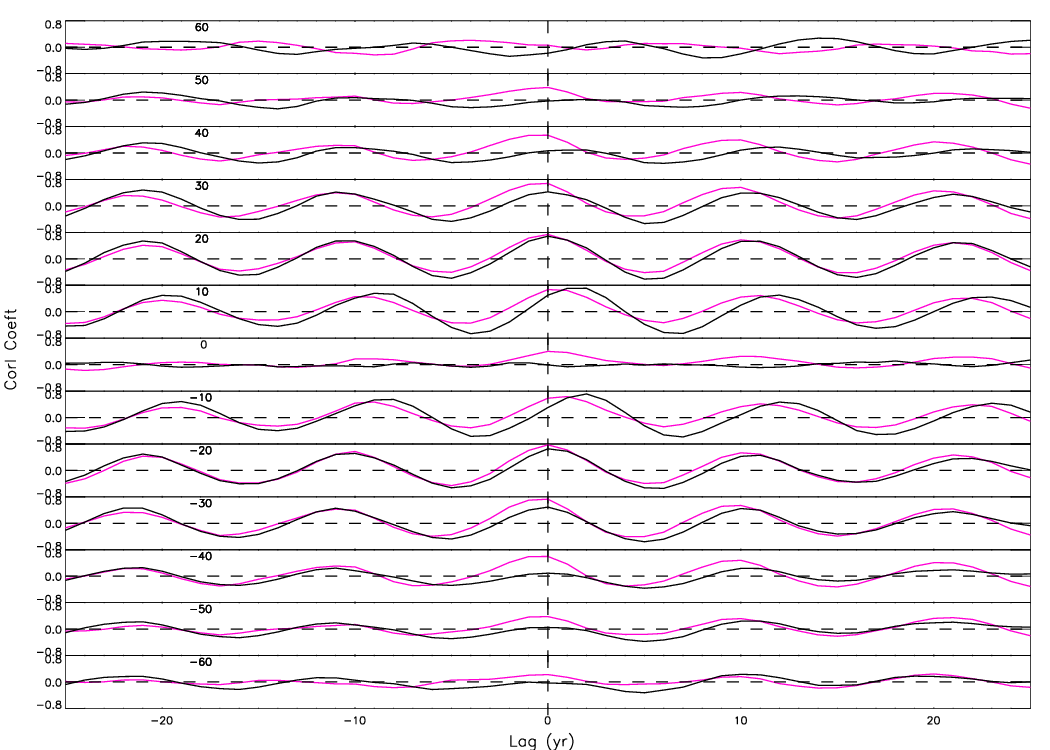}} 
\centering
{\includegraphics[width=5cm,height=12cm]{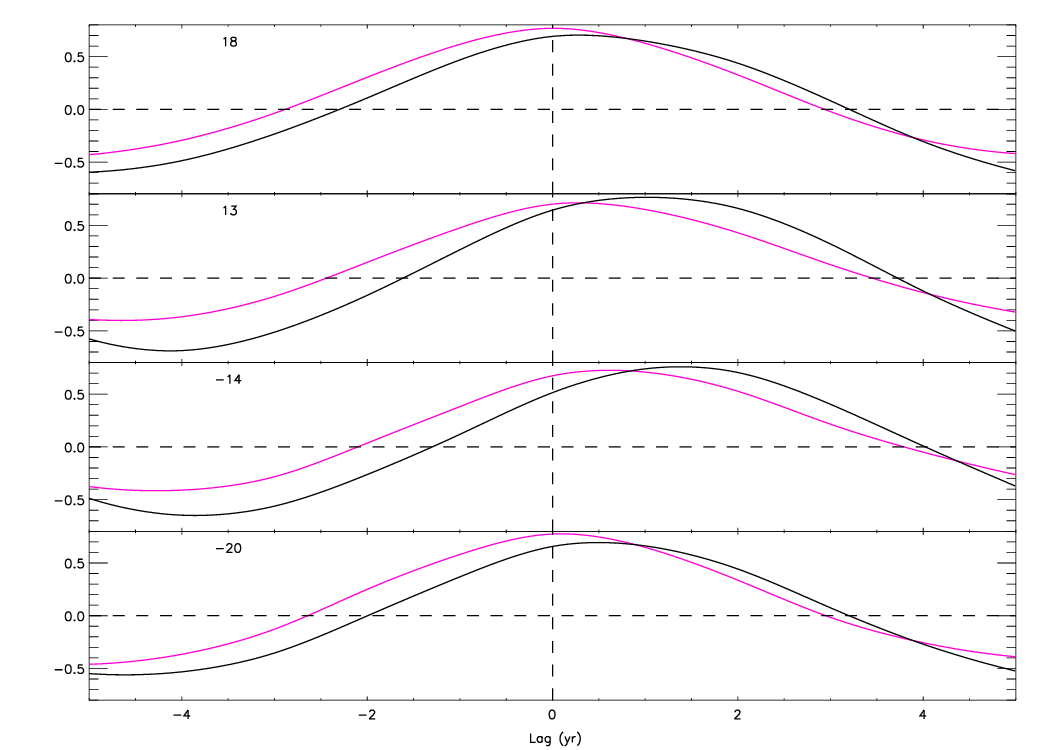}}
\caption{Cross-correlation of lane widths (pink) and intensities (black) with sunspot cycle as a function of lag (yr) at different latitudes. The panels on the right give a magnified view of the central region (-5 $<$ lag $<$ 5) at selected latitudes.]}
\label{fig:general}
\end{figure}

\section{Results} \label{sec:results}

We obtained the temporal variations of lane widths and intensities averaged over a year
in the latitude range 60$^{\circ}$ N to 60$^{\circ}$ S with an interval of 1$^{\circ}$. The variation is plotted in Figure 1 for 13 representative latitudes. The yearly averaged sunspot number variation is also shown. 
It is found that the latitudes near $\pm$20$^{\circ}$ 
have the solar cycle modulations. The lane width data generally have more
prominent modulations than the intensity data. The solar cycle variation of Ca K intensity is up to 5 \%.

Next, the cross-correlation of lane  width and intensity with sunspot cycle as a 
function of lag (yr) is obtained at the different latitudes. This is shown in 
Figure 2. As in Figure 1, the correlation of lane  width and intensity with sunspot cycle is very high in the latitudes near $\pm$ (11--22)$^{\circ}$. The maximum correlation coefficient is about 0.8 and it decreases towards higher and lower latitudes.
At 60$^{\circ}$ N and also at the equator, the lane width and intensity correlation curves are nearly in opposite phases with each other although with reduced correlation coefficients. Similar anticorrelations are reported in the polar activity indices by \citet{2023ApJ...944..218P}. A closer examination of latitudes shows significant differences in the behavior of lane widths and intensities. This is shown in the right panels of Figure 2.
The maximum correlation for the lane widths occurs 18$^{\circ}$ N and 20$^{\circ}$ S with no phase lag. For intensities, this happens at 13$^{\circ}$ N and 14$^{\circ}$ S with a phase lag of 1.25 to 1.5 yr. The results imply that there is no unique latitude that follows the sunspot cycle; the intensity and lane width follow the sunspot cycle in slightly different ways.

It can also be noted that there is a systematic change in 
the lag in latitudes. For both lane widths and intensities, the lag is zero near $\pm$ 20$^{\circ}$ which decreases towards higher
latitudes and increases towards the equator. There is a difference in the magnitude of lag for lane  width and intensity with the sunspot cycle. This aspect is studied in more detail in the following.

Note that the x-axis interval is 1 year. We aim to get the exact time lag
corresponding to the maximum correlation coefficient within the fraction of an
yr. For this, we took the data from Figure 2 in the range $\pm$5 yr (close to one
solar cycle) and interpolated it with an interval of 0.01 yr.
The maximum correlation coefficient of lane width and intensity with sunspot cycle at different latitudes are plotted in the top panel of Figure 3. 
A 5-point smoothing average is also plotted which shows the peaks
somewhat better. The peaks are still broader but can be specified with an
uncertainty of ± 2$^{\circ}$
It can be seen that the plots are nearly symmetrical with respect to the equator, but there are notable deviations from symmetry. The peaks are
closer to the equator in the North. The multiple components are seen at lower
levels in the North. There is also a sharp turnaround in the intensity correlations
beyond 50$^{\circ}$ N, although with reduced significance.
For lane width, the correlation coefficient is 0.45 at the equator, 0.8 near $\pm$ 20$^{\circ}$, and about 0.2 near $\pm$ 60$^{\circ}$. For intensity, correlation coefficient is close to zero at the equator, 0.8 near $\pm$ 13$^{\circ}$, and close to zero near $\pm$ 50$^{\circ}$. The correlation coefficients show a multi-component structure in both hemispheres. There is a marked phase  lag between the lane width and intensity plots; intensity reaches the maximum correlation in much lower latitudes in the two hemispheres. 
At the level of the correlation coefficient
= 0.5, the width of the lane width peaks is about 40$^{\circ}$, whereas the width of the
intensity peaks is 20$^{\circ}$.

The bottom panel shows the lag (yr) of lane width and intensity with solar cycle  at different latitudes. Only those points with a correlation coefficient above 0.29 -- highly significant correlation where the probability of occurrence by chance is less than 1 \% -- are shown. The plots are nearly symmetric here too. The lag varies from -0.5 yr to 0.8 yr for lane width. For intensity, the lag varies from -0.3 yr to about 2.5 yr. 
Also, note that there is a broad reversal in the lag values for both quantities; at about$\pm$ (25$\pm$2)$^{\circ}$ for intensity and $\pm$ (40$\pm$2)$^{\circ}$ for lane width.

It may be noted that the autocorrelation function from Ca K image windows gives two length scales of
supergranulation, namely, the lane width and the size \citep{2023ApJ...959L..24R}. Our results indicate a highly significant correlation between lane width and
sunspot cycle in the latitude range 55$^{\circ}$ S–55$^{\circ}$ N. The relationship between lane width and
size is not clear now. So, the current work may not resolve the conflicting reports on the
cycle dependency of the size, but it may give some useful, related information.

It is interesting to note that the intensities and the lane widths behave in different ways at different latitudes.  We know that both lane widths and intensities are dependent on local magnetic flux, so there could be other factors which may influence their behavior. Lane width is a measure of magnetic flux at the supergranular cell boundaries \citep{1964ApJ...140.1120S}. Intensity, on the other hand, also depends on the temperature and chemical abundance. Measurements of sunspots over the solar latitudes show  a peak around $\pm$ 13--15$^{\circ}$. \citep{1994ASPC...68....1H}. This is consistent with our results which show that the intensity peaks at 
(13$\pm$2)$^{\circ}$ N and (14$\pm$2)$^{\circ}$ S.
For lane width, the peaks appear after about 5$^{\circ}$. Magnetic flux is mainly transported on the solar surface by differential rotation, granular and supergranular convection, and meridional flows. \citet{2014ApJ...792...22S} find evidence of meridional flow from their Ca {\sc ii} K line profile analysis. The study needs to be continued with more datasets to ascertain the reasons for the behavior of the two quantities which may bring out interesting aspects of flux transport on the solar surface. 

\begin{figure}
\includegraphics[width=15cm,height=18cm]{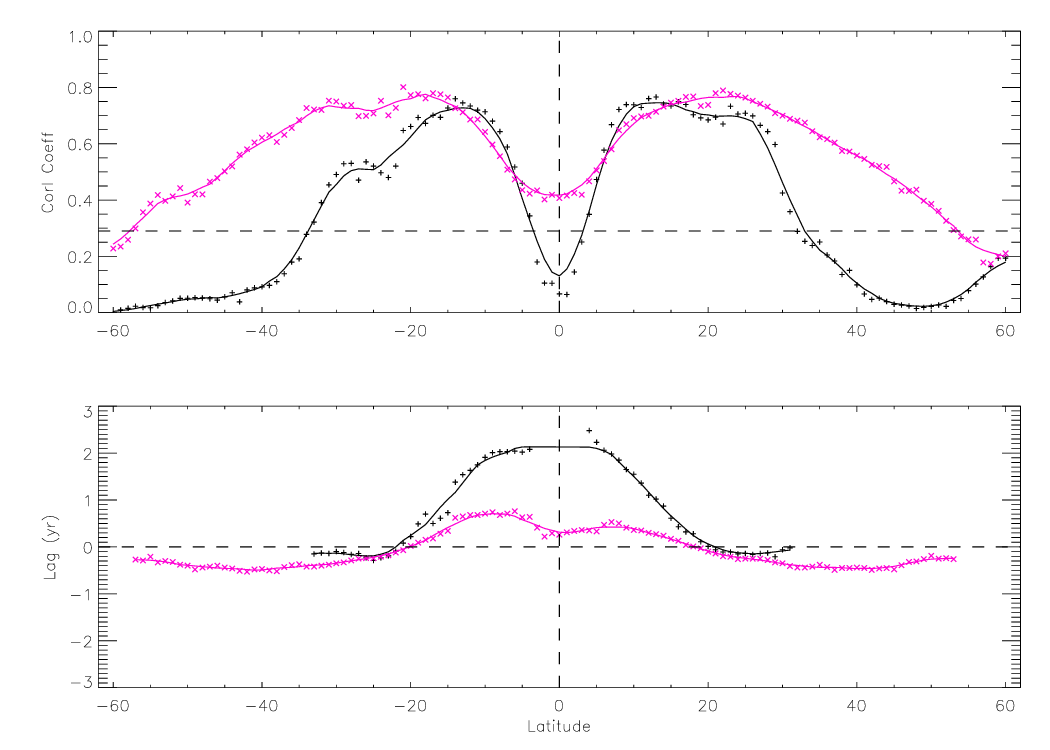}
\caption{Maximum correlation coefficient of lane  width (pink) and intensity (black) with sunspot cycle at different latitudes (top). The continuous line represents a 5-point smoothing average. The horizontal dashed line represents the 99 \% confidence level. Lag (yr) of lane width (pink) and intensity (black) with solar cycle  at different latitudes (bottom). Only those points with a correlation coefficient above 
0.29 (99 \% confidence level) are shown.
\label{fig:general}}
\end{figure}

\section{Conclusions} \label{sec:conclusions}

We conducted an extensive analysis of the newly calibrated Ca {\sc ii} K spectroheliograms from Kodaikanal archival data and obtained the intensities and supergranular lane widths as functions of latitude and time. We obtained the cross correlation function of these quantities with the sunspot cycle to see the variation of correlation coefficients as a function of time lag at each latitude within the range 60$^{\circ}$ N to 60$^{\circ}$ S  with an interval of 1$^{\circ}$. The following major points can be noted.

1. The main aim is to see whether any particular latitude follows the solar cycle exactly. The maximum correlation for the lane widths occurs at (18$\pm$2)$^{\circ}$ N and (20$\pm$2)$^{\circ}$ S. For intensities, this happens at (13$\pm$2)$^{\circ}$ N and (14$\pm$2)$^{\circ}$.
The results generally show that no unique latitude exactly follows the solar cycle for all quantities. For different physical quantities, the correlation reaches a maximum at different latitudes.

2. The above finding on maximum correlation is comparable to results of Howard (1994), which report that sunspot distribution over the solar latitudes peaks around $\pm$ 13--15$^{\circ}$. There is a good agreement for intensity, whereas there is a difference of 
about 5$^{\circ}$ for lane width.

3. The total solar irradiance depends on the sunspot cycle, and the amplitude of variation is about 0.1~\% from minimum to maximum \citep{2023JASTP.25206150C}. However, the variation of UV irradiance is more than an order of magnitude, about 1.5 \% \citep{1989Sci...244..197L}. For individual spectral lines, the variation can be even more. Our results show that the solar cycle variation of Ca K intensity is up to 5 \%.

4. Another interesting finding is that the lane width correlations peak during the solar maximum, while the intensity correlations peak 1.25--1.5 yr after the solar maximum.

5. The correlation coefficients of both quantities show an approximate North-South symmetry. The results show that for lane width, the correlation is significant in a 
broad range (about 55$^{\circ}$ N -- 55$^{\circ}$ S), whereas for intensity, the range is much narrower (about 5$^{\circ}$ -- 35$^{\circ}$ N and 5$^{\circ}$ -- 35$^{\circ}$ S). The lack of significant correlation for quiet Sun Ca K intensity at the solar equatorial region is 
unexpected.

6. Another important difference between lane width and intensity behavior is in the time lag. The lag varies from -0.5 yr to 0.8 yr for lane width; for intensity, it varies from -0.3 yr to about 2.5 yr. The reason for this significant difference is not apparent.

7. The current work may not resolve the controversy of the
cycle dependency of the supergranular size. Still, it can be concluded that the supergranular lane width depends on the solar cycle 
for the quiet Sun. A highly significant positive correlation is found in the latitude range 55$^{\circ}$ S –- 55$^{\circ}$ N.

Supergranular diffusion is one of the ways by which magnetic flux is transported on the solar surface. The phase lag between lane width and solar cycle is due to the flux transport, and its speed can be estimated. The speed of flux transport plays a vital role in the length of the solar cycle. 
Some of the above points (3,4, \& 5) are important to the quiet Sun UV irradiance variations. The implications of the results for the origin of supergranulation are to be understood through future studies.

\begin{acknowledgments}
This work is funded by the Department of Science and Technology, Government of India. We thank the numerous observers and the digitization team of Kodaikanal Solar Observatory for the availability of the data.  
\end{acknowledgments}

\bibliographystyle{aasjournal}

\end{document}